\documentclass[superscriptaddress,
 amsmath,amssymb,
 aps,
pra,
reprint
]{revtex4-2}

\usepackage{hyperref}
\usepackage{graphicx}
\usepackage{dcolumn}
\usepackage{bm}
\usepackage{physics}
\usepackage{amsfonts, amsmath}
\usepackage{dsfont}
\usepackage{amsthm}
\usepackage{bm}
\usepackage{bbold}
\usepackage{color}
\usepackage{lipsum,babel}
\usepackage[normalem]{ulem}
\usepackage{tabularx}

\newcommand{\cor}[1]{{\color{red}{#1}}}

\newcommand{\haf}{\text{haf}}

\newcommand{\per}{\text{Per}}
\newcommand{\poly}{\text{poly}}

\begin{document}
\definecolor{navy}{RGB}{46,72,102}
\definecolor{pink}{RGB}{219,48,122}
\definecolor{grey}{RGB}{184,184,184}
\definecolor{yellow}{RGB}{255,192,0}
\definecolor{grey1}{RGB}{217,217,217}
\definecolor{grey2}{RGB}{166,166,166}
\definecolor{grey3}{RGB}{89,89,89}
\definecolor{red}{RGB}{255,0,0}

\preprint{APS/123-QED}

\title{Quantum-inspired classical algorithm for graph problems by Gaussian boson sampling}
\author{Changhun Oh}
\email{changhun@uchicago.edu}
\affiliation{Pritzker School of Molecular Engineering, University of Chicago, Chicago, Illinois 60637, USA}
\author{Bill Fefferman}
\affiliation{Department of Computer Science, University of Chicago, Chicago, Illinois 60637, USA}
\author{Liang Jiang}
\affiliation{Pritzker School of Molecular Engineering, University of Chicago, Chicago, Illinois 60637, USA}
\author{Nicolás Quesada}
\affiliation{Department of Engineering Physics, {\'{E}}cole Polytechnique de Montr{\'{e}}al, Montr{\'{e}}al, QC, H3T 1JK, Canada}

\begin{abstract}
We present a quantum-inspired classical algorithm that can be used for graph-theoretical problems, such as finding the densest $k$-subgraph and finding the maximum weight clique, which are proposed as applications of a Gaussian boson sampler.
The main observation from Gaussian boson samplers is that a given graph's adjacency matrix to be encoded in a Gaussian boson sampler is nonnegative, which does not necessitate quantum interference.
We first provide how to program a given graph problem into our efficient classical algorithm.
We then numerically compare the performance of ideal and lossy Gaussian boson samplers, our quantum-inspired classical sampler, and the uniform sampler for finding the densest $k$-subgraph and finding the maximum weight clique and show that the advantage from Gaussian boson samplers is not significant in general.
We finally discuss the potential advantage of a Gaussian boson sampler over the proposed classical sampler.
\end{abstract}

\maketitle

\section{Introduction}
Over the last few years, we have seen the first plausible quantum computational advantages from random circuit sampling with superconducting qubits \cite{arute2019quantum, wu2021strong, bouland2019complexity, boixo2018characterizing} and Gaussian boson sampling \cite{zhong2020quantum, zhong2021phase, madsen2022quantum, hamilton2017gaussian, deshpande2022quantum}.
While there are numerous interesting debates on the claimed quantum advantage, such as the effect of noise and the verification \cite{gao2021limitations, bouland2022noise, oh2022spoofing, aharonov2022polynomial, oh2023classical}, the computational cost of classically simulating the existing quantum devices is still enormous.
Therefore, an obvious next step beyond proof-of-principle experiments is to take advantage of the computational power of such noisy intermediate-scale quantum (NISQ) devices to solve more practical problems.

An interesting and crucial observation to exploit the potential quantum advantage of Gaussian boson sampling is that one can embed a graph in the circuit so that a Gaussian boson sampler can be programmed for various graph-theoretical problems~\cite{bradler2018gaussian, arrazola2018using, banchi2020molecular, bromley2020applications}.
Such an observation has led many theoretical proposals to solve problems \cite{bromley2020applications}, such as finding the densest $k$-subgraph \cite{arrazola2018using} and finding the maximum weight clique \cite{banchi2020molecular}.
These problems have attracted much attention because of their potential applications in a wide range of fields such as data mining \cite{kumar1999trawling, boginski2006mining, 10.14778/2168651.2168658, balasundaram2011clique, pattillo2011clique} and bioinformatics \cite{fratkin2006motifcut, saha2010dense, malod2010maximum} (see Ref.~\cite{bromley2020applications} for more applications).
Due to the importance of solving the problems, such theoretical proposals have recently started to be experimentally implemented \cite{banchi2020molecular, yu2022universal, sempere2022experimentally}, which opens the possibility of taking advantage of quantum computational advantage from NISQ devices to solve practical problems.
In particular, the quantum advantage was claimed by comparing it with the uniform distribution, which was referred to as the classical algorithm.

Meanwhile, potential applications of quantum devices sometimes turn out to be classically simulable.
For example, the molecular vibronic spectra problem has been considered as an application of Gaussian boson sampling \cite{huh2015boson}.
Very recently, classical algorithms have been developed to solve the problem as accurately as the Gaussian boson sampler for many cases, including the Fock-state version of the problem~\cite{oh2022quantum}.
Therefore, to claim and exploit the potential quantum advantage more rigorously, it is also crucial to scrutinize the problems' complexity and potential ways of simulating the quantum algorithm using a classical counterpart.
Similarly, while the graph-theoretical problems we consider, such as finding the densest $k$-subgraph and finding the maximum weight clique, have been considered applications of Gaussian boson sampling, it has been a longstanding open question whether they provide a provable quantum computational advantage \footnote{There was a claim that there exists an efficient classical algorithm to solve the problems (e.g., Refs.~\cite{aaronson2020, aaronson2023certified}), but to the best of our knowledge, the algorithm and its details have not been presented.}.

In this work, we present a quantum-inspired classical sampler that can be used for graph-theoretical problems, such as finding the densest $k$-subgraph and finding the maximum weight clique.
The algorithm is inspired by the proposals of using Gaussian boson samplers for the problems \cite{arrazola2018quantum, arrazola2018using, banchi2020molecular, bromley2020applications}.
The main idea behind exploiting a Gaussian boson sampler to solve such problems is that the density of a graph is typically proportional to the number of perfect matchings and that the output probability of a Gaussian boson sampler is proportional to the latter.
Therefore, a Gaussian boson sampler enables us to more frequently sample subgraphs with many perfect matchings, implying a high density.
Our proposed classical algorithm has the same property as a Gaussian boson sampler in that the sampler more frequently samples subgraphs with many perfect matchings.

The key observation that makes our classical algorithm efficient is that the adjacency matrix of a graph to be embedded in Gaussian boson sampling circuits is always nonnegative.
In addition, it is often the case that computational problems associated only with nonnegative quantities are easier than more general cases.
Using this observation, we present a method that embeds the adjacency matrix of a given graph into two-photon boson sampling circuits that can be classically efficiently simulated.
We emphasize that a Gaussian bosons sampler and our classical algorithm depend differently on the number of perfect matchings; the former is proportional to its square while the latter is to itself.
To see the effect of such a difference,
we then compare the performance of the Gaussian boson sampler, the quantum-inspired classical sampler, and the uniform sampler.
We consider Erdős–Rényi random graphs with different parameters and show that the average density and maximum density of subgraphs obtained by a Gaussian boson sampler and our classical sampler do not exhibit a significant difference.
We also compare the performance of a heuristic classical algorithm equipped with each sampler for finding the maximum weighted clique and show again that the difference between our classical sampler and the Gaussian boson sampler is not significant.
We finally discuss whether or not a Gaussian boson sampler has a potential exponential advantage over our quantum-inspired classical algorithm.

The paper is organized as follows:
In Sec.~\ref{sec:GBS}, we review the relation between Gaussian boson sampling and graph-theoretic problems.
In Sec.~\ref{sec:classical}, we provide our quantum-inspired classical algorithm and how to program graph-theoretic problems to the algorithm.
In Sec.~\ref{sec:performance}, we numerically analyze the algorithm's performance by comparing it with the Gaussian boson sampler and the uniform sampler.
In Sec.~\ref{sec:advantage}, we discuss the potential advantage of Gaussian boson sampler over our classical algorithm.

\section{Gaussian boson sampling and its application to graph-theoretic problems}\label{sec:GBS}
Gaussian boson sampling is a sampling task which is proven to be hard to classically simulate under plausible assumptions \cite{hamilton2017gaussian, deshpande2022quantum, grier2022complexity}.
It can be experimentally implemented by injecting squeezed states with squeezing parameters $\{r_i\}_{i=1}^M$ into $M$ linear-optical circuit, characterized by an $M\times M$ unitary matrix, and measuring the photon number distribution over the output modes.
The output probability of obtaining the photon-number outcome $\bm{n}\in\mathbb{Z}_{\geq 0}^M$ can be expressed by the hafnian of a relevant matrix $A$ as \cite{hamilton2017gaussian}
\begin{align}\label{eq:prob}
    p(\bm{n})=\frac{|\haf{A}_{\bm{n}}|^2}{\bm{n}!\sqrt{|\Sigma+\mathbb{1}/2|}},
\end{align}
where $\Sigma$ is the covariance matrix of the output quantum state \cite{weedbrook2012gaussian, serafini2017quantum} and $A=UDU^\text{T}$ with a diagonal matrix $D=\text{diag}(\{\tanh r_i\}_{i=1}^M)$, and $A_{\bm{n}}$ is obtained by selecting rows and columns of matrix $A$ corresponding to the outcome $\bm{n}$.
Here, the hafnian of a $2n\times 2n$ matrix $X$ is defined as
\begin{align}
    \haf(X)=\frac{1}{2^n n!}\sum_{\sigma \in\mathcal{S}_{2n}}\prod_{i=1}^n X_{\sigma(2i-1),\sigma(2i)},
\end{align}
where $\mathcal{S}_{2n}$ is the $2n$-element permutation group.
The key idea of using Gaussian boson sampling for various applications is that one can embed an arbitrary complex symmetric matrix $A$ by using the Takagi decomposition of $A=UDU^\text{T}$ with an appropriate rescaling to make the diagonal component less than one, which is to assure that $\tanh r_i$'s are within their range \cite{bradler2018gaussian}.
We note that there is freedom of choosing the rescaling factor as long as $\tanh r_i$'s are smaller than one, which changes the total photon number distributions but not the relative weight in the same total photon number outcomes.
Hence, for a given adjacency matrix $A$, one constructs a Gaussian boson sampling circuit with squeezing parameters obtained by the diagonal matrix $D$ and a linear-optical circuit with the unitary matrix $U$.


Most of the proposed applications are related to graph-theoretic problems due to the hafnian and its relation to the output probability of Gaussian boson sampling \cite{bromley2020applications, oh2022classical}.
Consider an undirected graph $G=(V,E)$, where $V$ is the set of vertices and $E\subset V\times V$ is the set of edges.
The graph can be represented by its $|V|$ dimensional adjacency matrix $A$ whose matrix element $A_{ij}=1$ when $(i,j)\in E$, i.e. when there is an edge between the $i$th and $j$th vertices.
The key observation is that the hafnian of a matrix $A$ is the number of perfect matchings of a graph whose adjacency matrix is $A$.
Therefore, by embedding the adjacency matrix $A$ into Gaussian boson sampling, one can obtain a sampler that favorably samples outcomes corresponding to a subgraph whose number of perfect matching is large.
Such a property has been used to apply a Gaussian boson sampling for finding dense subgraphs based on the observation that the density of a graph is likely to be large when its number of perfect matchings is large \cite{arrazola2018using}.
Hence, it can lead to an acceleration of many heuristic classical algorithms which employ sampling a subgraph as their subroutines, such as finding the densest $k$-subgraph problem \cite{arrazola2018using} or finding the maximum clique \cite{banchi2020molecular}.

In the following section, we will show that a similar acceleration is possible using a classical algorithm.
The critical observation is that the adjacency matrix $A$ does not contain negative elements and that many computational problems restricted to nonnegative elements are often easier than more general cases.
For example, computing the permanent of nonnegative-element matrices in a multiplicative error is easy for a classical computer \cite{jerrum2004polynomial} while the corresponding problem for general matrices is hard (\#P-hard) \cite{jerrum2004polynomial}.
Our intuition from a physical perspective is that Gaussian boson sampling associated with nonnegative-element matrix $A$ does not necessarily require a complicated multi-photon interference.
Indeed, this is true for Fock-state boson sampling; if we use distinguishable particles as an input of Fock-state boson sampling, which is obviously easy to simulate instead of indistinguishable bosons, the output probability is expressed by the permanent of nonnegative-element matrices \cite{aaronson2011computational}.


\begin{figure*}[t]
\includegraphics[width=350px]{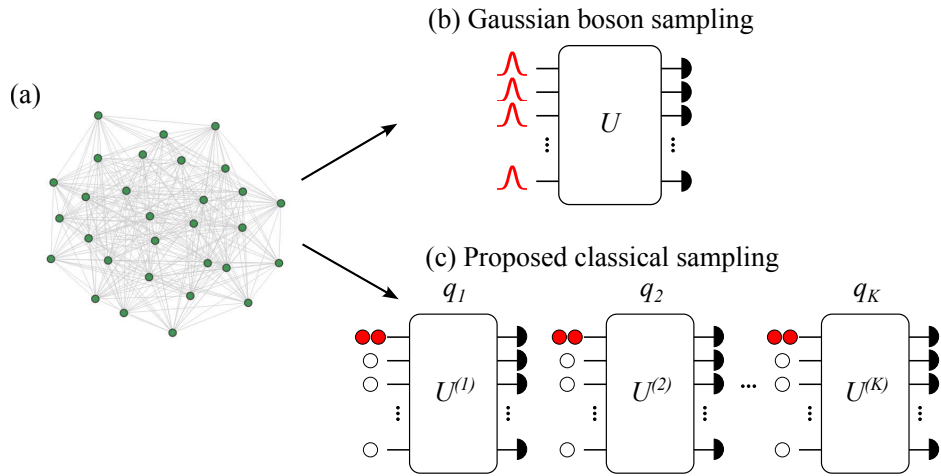}
\caption{(a) Graph related to a given graph-theoretical problem such as finding the densest $k$-subgraph or the maximum weighted clique. (b) One can embed such a graph into a Gaussian boson sampler \cite{bradler2018gaussian} (c) We propose a classical sampler which can solve the problem in a similar manner to a Gaussian boson sampler.}
\label{fig:scheme}
\end{figure*}

\section{Classical sampling algorithm for graph-theoretic problems}\label{sec:classical}
\subsection{Multiple two-photon boson sampling}
Before we provide our classical algorithm, let us consider a Fock-state boson sampling with input $|20\dots 0\rangle$, i.e., two-photon boson sampling.
After a linear-optical circuit $U$, the probability of obtaining photons at the $i$th and $j$th modes with $i\neq j$ is given by
\begin{align}
    p_{ij}=\frac{|\per U_{1,(i,j)}|^2}{2}=2|U_{1i}|^2|U_{1j}|^2,
\end{align}
where $U_{1,(i,j)}$ is the submatrix of $U$ selecting the first rows twice and the $i$th and $j$th columns.
When $i=j$, $p_{ii}=|U_{1i}|^4$.
More simply, the probability $p_{ij}$ can be thought of as that of obtaining two photons at $i$ and $j$'s modes after two independent trials with probability $p_i=|U_{1i}|^2$.

Now, consider $K$ different two-photon boson sampling circuits with different linear-optical circuits $\{U^{(k)}\}_{k=1}^K$.
We will denote $p_{ij}^{(k)}$ as the $k$th circuit's probability of obtaining photons at the $i$th and $j$th modes.
Let us consider a sampling such that for $N$ trials, we randomly choose one circuit out of the set of circuits with probability $q_k$ with $\sum_{k=1}^Kq_k=1$ and inject two photons in the first mode.
Finally, we always obtain a $2N$ number of photons after finishing $N$ trials.
Let us compute the probability of obtaining outcome $\bm{n}=(1,\dots,1,0,\dots,0)$, i.e., $2N$ photons are detected on the first $2N$ modes.
One can intuitively see that this is related to the perfect matchings, i.e. to hafnian, because we need to find matchings of photon pairs that originate from the same circuit.
More precisely and formally, the probability is the sum of possible configurations that provide the corresponding output:
\begin{align}
    p(\bm{n})
    &=\frac{1}{2^N}\sum_{\sigma\in \mathcal{S}_{2N}}\prod_{i=1}^N\sum_{k=1}^K q_{k}p^{(k)}_{\sigma(2i-1),\sigma(2i)} \\ 
    &=\sum_{\sigma\in \mathcal{S}_{2N}}\prod_{i=1}^N\sum_{k=1}^K q_{k}|U_{1,\sigma(2i-1)}^{(k)}|^2|U_{1,\sigma(2i)}^{(k)}|^2 \\ 
    &=2^N N!\haf(A_{\bm{n}}),
\end{align}
where $A$ is a nonnegative $M\times M$ matrix with its elements defined as
\begin{align}\label{eq:A_form}
    A_{ij}\equiv \sum_{k=1}^K q_{k}|U_{1,i}^{(k)}|^2|U_{1,j}^{(k)}|^2=VQV^\text{T}=WW^\text{T},
\end{align}
and
\begin{align}
    V_{ik}\equiv |U_{1,i}^{(k)}|^2, ~ W_{ik}\equiv \sqrt{q_k}|U_{1,i}^{(k)}|^2,~ Q\equiv \text{diag}(q_1,\dots,q_K).
\end{align}
Also, $A_{\bm{n}}$ is a $2N\times 2N$ submatrix of $A$ with selecting the part of ones of $\bm{n}$.
The expression of the probability suggests that if we implement the routine as described above, the corresponding sampler's output probability is proportional to the hafnian of the submatrix of $A$.
Thus, such a construction allows us to find a classical sampler whose output probability is written as a submatrix of the hafnian of a matrix $A$ which can be obtained by $U$ and $Q$.
The remaining challenge is to prove that we can embed an arbitrary nonnegative matrix $A$ into a sampler whose output probability is proportional to the hafnian of submatrices of $A$, which will be addressed in the following section.


\subsection{Mapping a general graph to a circuit}
Consider an $M\times M$ general nonnegative symmetric matrix $A$, which corresponds to a graph of $M$ vertices.
Although we will call this matrix an adjacency matrix for simplicity, we only require the matrix to be nonnegative and symmetric.
We first note that the hafnian of a matrix does not depend on its diagonal elements.
Thus, we can freely add an arbitrary diagonal matrix without changing its hafnian so that the matrix becomes diagonally dominant, i.e., $A_{ii}\geq \sum_{j\neq i}|A_{ij}|$ for all $i$'s.
Hence, without loss of generality, we will assume that a given adjacency matrix $A$ satisfies 
\begin{align}\label{eq:diag}
    A_{ii}= \sum_{j\neq i}|A_{ij}|~~~~ \text{for all $i$'s}.
\end{align}
It is known that a nonnegative symmetric diagonally dominant matrix is completely positive \cite{berman2003completely}, i.e., $A=HH^\text{T}$, where $H$ is not necessarily a square matrix.
Thus, once we find the matrix $H$, we can construct $W$ in Eq.~\eqref{eq:A_form}.
Now we show how to find $H$ such that $A=HH^\text{T}$, which is the result from Ref.~\cite{berman2003completely}.

For a given $M\times M$ adjacency matrix, after adding the diagonal matrix to satisfy Eq.~\eqref{eq:diag}, we can always decompose the matrix $A$ as
\begin{align}\label{eq:AB}
    A=\sum_{1\leq j<i\leq M}B^{(i,j)},
\end{align}
where the $M\times M$ matrices $B^{(i,j)}$ have $A_{ij}$ in positions $ii$, $ij$, $ji$ and $jj$ and 0 elsewhere.
Then, we can rewrite it as $B^{(i,j)}=\bm{b}^{(i,j)}(\bm{b}^{(i,j)})^\text{T}$ with the $M$ dimensional vector $\bm{b}^{(i,j)}$ whose $i$th and $j$th elements are $\sqrt{A_{ij}}$.
We then construct $M\times M^2$ matrix $H$ such that $H$'s $(M(i-1)+j)$th column is $\bm{b}^{(i,j)}$. 
Here, we set $(M(i-1)+i)$th columns to be zero.
Then, we can easily check that $A=HH^\text{T}$.

\begin{figure*}[t]
\includegraphics[width=500px]{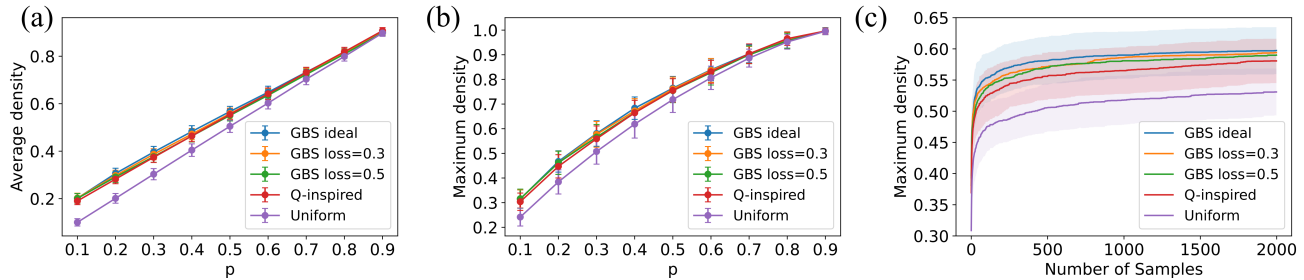} 
\caption{Density of samples from ideal and lossy Gaussian boson sampler, our classical sampler, and the uniform sampler for random graphs with $n=30$ vertices and $k=10$ size subgraphs and various $p$. (a) Average density for different regimes. (b) Maximum density for different regimes. We use $10^3$ different graphs with $10^2$ number of samples for each $p$. (c) The maximum density achieved by a different number of samples with $p=0.3$.}
\label{fig:density}
\end{figure*}

So far, we have shown how to find $H$, such that $A=HH^\text{T}$.
Now we will show how to construct two-photon boson sampling circuits $\{U^{(k)}\}_{k=1}^{M^2}$ to program $H$.
We first define an $M^2\times M^2$ diagonal matrix $D$ such that $D$'s each element is the sum of $H$'s $i$th column i.e., $D_{ii}=\sum_{k=1}^{M}H_{ki}$ for all $i\in [M^2]$.
We then divide $H$'s $i$th column by it, so that the resultant $M\times M^2$ matrix $V$ satisfies $\sum_{k=1}^M V_{ki}=1$ for all $i$'s and $HH^\text{T}=VD^2V^\text{T}$.
Finally, we rescale the matrix $D^2$ so that its trace becomes 1, i.e., $Q\equiv D^2/\Tr[D^2]$.
Finally, we obtain the same form as Eq.~\eqref{eq:A_form},
\begin{align}
    A=\text{Tr}[D^2]VQV^\text{T},
\end{align}
with a coefficient $\text{Tr}[D^2]$.
Here, the meaning of the diagonal matrix $Q$ is manifest that $Q$'s $i$th diagonal element is $q_k$, i.e., the probability of selecting the $k$th circuit.
Also, $V$'s $k$th column represents $|U_{1,i}^{(k)}|^2$.
Thus, accordingly, for each $k\in[M^2]$ we construct $U_{1,i}^{(k)}$, which can be easily done because we only require a condition for a single row of the unitary matrix and $V$ is a nonnegative matrix.
(This is to show the correspondence; in practice, one can directly use $|U_{1,i}^{(k)}|^2$ as sampling probabilities without constructing two-photon boson sampling.)

To summarize, we have shown that for a given $M\times M$ adjacency matrix $A$, we can construct a set of linear-optical circuits $\{U^{k}\}_{k=1}^{M^2}$ with associated probabilities $\{q_k\}_{k=1}^{M^2}$, which samples outcomes with probabilities proportional to the hafnian of the corresponding $2N\times 2N$ submatrix of $A$, i.e. of subgraphs of $2N$ vertices.
Here, we can freely choose $N$ depending on the size of subgraphs we want.
Since two-photon boson sampling can be easily simulated by a classical computer, the entire procedure can be simulated using a classical computer efficiently.

One interesting difference between our classical algorithm and a Gaussian boson sampler is that the latter's output probability is proportional to the square of the hafnian, $\propto |\haf(A_{\bm{n}})|^2$ instead of the hafnian itself $\propto \haf(A_{\bm{n}})$, which may provide better performance (see Sec.~\ref{sec:advantage} for more discussion).
In the following section, we compare their performances for finding the densest $k$-subgraph problem and finding the maximum clique problem.
One advantage of our quantum-inspired classical algorithm is that we can choose the output photon number as we want, while Gaussian boson sampling has a fixed total photon number distribution for a given setup.
Thus, when one wants to focus on a particular size of subgraphs, Gaussian boson samplers cost additional overhead for post-selection.

We emphasize that since we require the matrix $A$ to be nonnegative, our algorithm cannot be used for the standard Gaussian boson sampling where the matrix $A$ is generally complex.
Also, for the relevant problems we consider, we only need to consider collision-free outcomes, since only collision-free outcomes corresponds to proper subgraphs.
Due to this and since the constructed $U_{1,i}^{(k)}$ is only nonzero for at most two $i$'s for each $k$ (see Eq.~\eqref{eq:AB} and the construction of $B$, $H$, and $V$), we only need to generate two photons for such $i$'s modes instead of sampling once we sample a circuit from $\{q_k\}_{k=1}^{M^2}$.
Finally, our sampling algorithm can also be used to estimate the hafnian of a nonnegative matrix with an additive error, which is comparable with the algorithms in \cite{barvinok1999polynomial, rudelson2016hafnians}.

\section{Performance comparison}\label{sec:performance}
Many applications of the Gaussian boson sampling routine are based on generating subgraphs that have high density.
Typically, the performance of Gaussian boson sampling has been compared to the uniform sampler, which was treated as a representative classical sampler \cite{arrazola2018using, sempere2022experimentally, banchi2020molecular, yu2022universal}.
Our quantum-inspired algorithm clearly provides a better way of exploiting a classical computer than a simple uniform sampler.
In this section, we quantitatively compare the performances for two different tasks.

\subsection{Finding the densest $k$-subgraph}
One of the particularly interesting problems proposed to use Gaussian boson samplers is finding the densest $k$-subgraph problem, the definition of which is as follows:
Given a graph $G$ with $M$ vertices, find the subgraph of $k<M$ vertices with the largest density.
This problem is known to be NP-hard \cite{feige2001dense}.
Thus, generally, it is not believed to be efficiently solved using quantum devices. 
Nevertheless, in Ref.~\cite{arrazola2018using}, it was shown that a Gaussian boson sampler can accelerate the performance of a heuristic classical algorithm using the fact that the density of a graph is proportional to the number of perfect matchings and that the output probability of a Gaussian boson sampler is proportional to the square of the number of perfect matchings.
As we have shown in the previous section, our classical sampler also enjoys the same property, namely, its output probability is proportional to the number of perfect matchings.
Therefore, the key property of exploiting a Gaussian boson sampler to solve this problem can also be obtained using our classical algorithm.

For the simulation with the Gaussian boson sampler, for a given graph, we program the Gaussian boson sampling circuit using the method in Sec.~\ref{sec:GBS} and exactly simulate it using a classical algorithm \cite{gupt2019walrus}.
Here, we choose the covariance matrix whose average photon number is equal to the target subgraph's size $k$.
As mentioned before, for our quantum-inspired classical algorithm, we can fix the output photon number, corresponding to the size of subgraphs.
The uniform sampler is also implemented over the $k$ photon subspace.


For practical consideration, we also consider lossy Gaussian boson sampling, where we replace the lossless squeezed states with squeezing parameter $r_i$ by new squeezing parameter $r_i'$ such that the average photon number of the new squeezed states after loss equals the mean photon number of the lossless squeezed states, i.e.,
$\sinh^2 r_i = \eta \sinh^2 r_i'$ where $\eta$ is the transmission rate and $1-\eta$ is the loss rate.
Thus, we compensate for the effect by increasing the input squeezing parameter so that the output Gaussian state has the average photon number to be $k$.
Otherwise, the probability of detecting $k$ photons is highly reduced due to the loss.
From the experimental perspective, such an adjustment inevitably increases the input squeezing parameter, making the experiment more demanding.

\begin{figure}[t]
\includegraphics[width=250px]{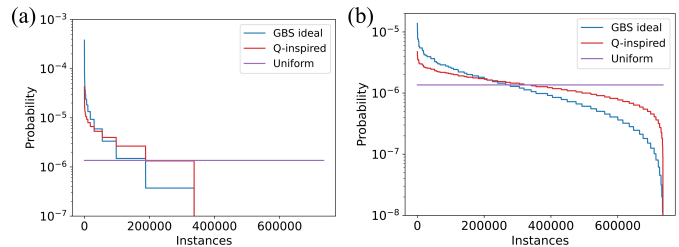} 
\caption{Examples of probability distributions of samplers for Erdős–Rényi random graphs with (a) $p=0.3$ and (b) $p=0.75$ with $n=24$, $k=8$. The ratio of the largest probability between the Gaussian boson sampler and the uniform sampler is about $280$ and $10$ , while that between the Gaussian boson sampler and the quantum-inspired classical algorithm is only about $9$ and $3$, respectively.}
\label{fig:density_prob}
\end{figure}

To compare the performance between different strategies, the ideal and lossy Gaussian boson samplers, our quantum-inspired classical sampler, and the uniform sampler, we generate Erdős–Rényi random graphs with different probability $p$ of having edges between vertices and present the behavior of the density of generated samples from each sampler in Fig.~\ref{fig:density}.
The figure shows that the Gaussian boson sampler and our quantum-inspired classical sampler do not have a significant difference for a wide range of $p$, while the uniform sampler tends to generate lower density graphs than the former two, especially for low $p$.
We also compare the maximum density obtained for a different number of samples.
Again, the uniform sampler clearly shows poor performance than other samplers.
Although our quantum-inspired classical algorithm generates a slightly lower maximum density, the difference is not very significant.

To compare lossy Gaussian boson samplers with the lossless Gaussian boson sampler, we observe that the performance degradation due to loss is insignificant.
Hence, for the task of generating dense subgraphs, adjusting the input squeezing parameters maintains the performance of the ideal Gaussian boson sampler, while preparing squeezed states with large squeezing becomes more demanding experimentally and may cause additional loss.
Recently, such noise robustness has been used to provide another classical way of generating dense graphs in Ref.~\cite{solomons2023gaussian}.

In Fig.~\ref{fig:density_prob}, we compare the probability distributions of the ideal Gaussian sampler (normalized to the post-selected sector), the quantum-inspired classical sampler, and the uniform sampler for Erdős–Rényi random graphs with $p=0.3$ and $p=0.75$ cases.
For the instance that has the largest probability, i.e., the largest perfect matching, the probability ratio between the Gaussian boson sampler and the uniform sampler is large as $280$ and $10$, respectively, which can be a large advantage.
However, when compared to our quantum-inspired classical algorithm, the ratio becomes only about $9$ and $3$.
It implies that we only need about $\sim 10$ times more samples to obtain the instance.
Such a difference comes from the different proportionality of the probabilities to hafnian.
We discuss this difference and the potential advantage of Gaussian boson samplers over our algorithm in Sec.~\ref{sec:advantage}.
In Refs.~\cite{arrazola2018using, arrazola2018quantum}, the additional heuristic classical algorithm is applied to find the densest $k$-subgraph.
In the next section for finding the maximum weighted clique, we will compare the performance incorporated with the additional algorithm.

\subsection{Finding the maximum weighted clique}
Another relevant interesting problem is finding the maximum weighted clique \cite{bomze1999maximum, wu2015review}.
In Ref.~\cite{banchi2020molecular}, the quantum-classical hybrid algorithm equipped with Gaussian boson samplers has been proposed to accelerate a heuristic classical algorithm to solve molecular docking problems, which is related to drug design \cite{kuntz1982geometric, kuhl1984combinatorial, banchi2020molecular}.
The problem takes an input of an undirected graph with vertices and its adjacency matrix with an additional vector, which is the weight vector associated with vertices.
Ref.~\cite{banchi2020molecular} provides a way of embedding the problem to a Gaussian boson sampler, which favorably samples the outcome corresponding to a high-weight clique.
To embed the weight of vertices in the sampler, we first construct a Gaussian boson sampler corresponding to a matrix $B$, which is given by
\begin{align}
    B=\Omega(D-A)\Omega,
\end{align}
where $\Omega$ is a suitable diagonal matrix and $D$ is degree matrix ($D-A$ is graph Laplacian).
The degree matrix is the diagonal matrix with the degree of each vertex.
Then, it can be shown that the probability distribution is given by
\begin{align}
    p(\bm{n})\propto [\det(\Omega_{\bm{n}})\haf(A_{\bm{n}})]^2.
\end{align}
Therefore, by choosing $\Omega$ as a diagonal matrix with the weight vector as its diagonal elements, we can introduce the effect of the weight in the probabilities, so that the sampler favors the outcomes having a large weight.
For collision-free cases, which is our main interest for graph problems, the dependence from $D$ disappears and we can set $B=\Omega A \Omega$ without loss of generality.

Using the fact that for $W=\text{diag}(w_1,\dots,w_n)$ \cite{barvinok2016combinatorics},
\begin{align}
    \haf(\Omega A\Omega)=\left(\prod_{i=1}^n w_i\right)\haf(A),
\end{align}
we can apply the same classical algorithm with the matrix $\Omega A \Omega$ because if $A$ is completely positive, then so is $\Omega A \Omega$.
As pointed out in Ref.~\cite{banchi2020molecular}, one might choose $\Omega$ as a weight matrix, but one can also choose $\Omega_{ii}=1+\alpha w_i$ to give some freedom that might weigh more on the hafnian than weight.
For simplicity, we choose $\alpha=1$ for our numerical results.

\begin{figure}[t]
\includegraphics[width=250px]{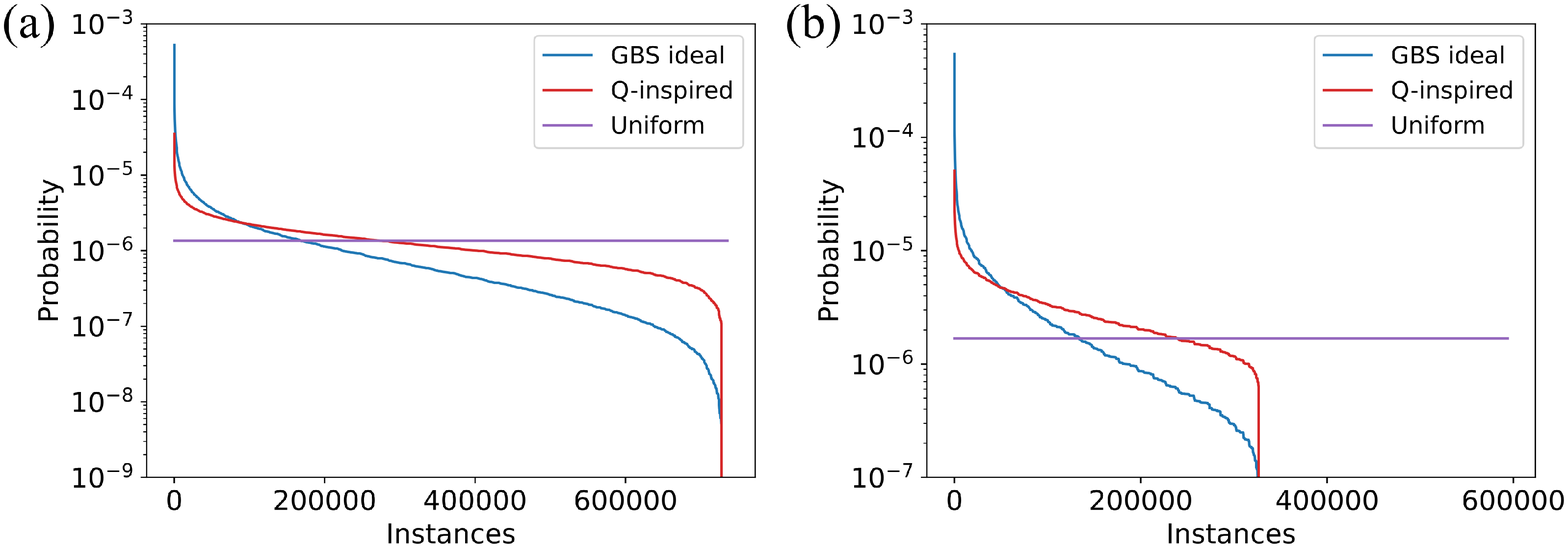} 
\caption{Probability distributions of samplers for (a) TACE-AS and (b) 1ow7, used in Ref.~\cite{banchi2020molecular}. The ratio of the largest probability between the Gaussian boson sampler and the uniform sampler is about $323$ and $11$, while that between the Gaussian boson sampler and the quantum-inspired classical algorithm is only about $385$ and $15$, respectively.}
\label{fig:clique_prob}
\end{figure}

We consider two graphs corresponding to different types of protein, which are studied in Ref.~\cite{banchi2020molecular}:
binding interaction between the tumor necrosis factor-$\alpha$ converting enzyme (TACE) and a thiol-containing aryl sulfonamide compound (AS) and
a different protein structure (PBD ID: 1ow7), corresponding to Paxillin LD4 motif bound to the Focal Adhesion Targeting (FAT) domain of the Focal Adhesion Kinase.

With the graphs and weights used in Ref.~\cite{banchi2020molecular}, we first obtain $10^5$ samples from each sampler and count the number of samples corresponding to the maximum weighted clique.
While the Gaussian boson sampler found $\sim 20$ number of samples corresponding to the maximum weighted clique for both cases, our quantum-inspired classical sampler and the uniform sampler found only one or two samples for both cases, which implies that Gaussian boson sampling indeed performs better than the other samplers.
The difference in the performance from the previous densest subgraph case is that the problems we consider for the maximum weight clique have only a single subgraph solution over many subgraphs. 
In contrast, the previous density problems may have many subgraphs having the same density.
To be more clear, in Fig.~\ref{fig:clique_prob}, we compare the probability distributions of the ideal Gaussian sampler, the quantum-inspired classical sampler, and the uniform sampler.
For the instance that has the largest probability, i.e., the largest perfect matching, the probability ratio between the Gaussian boson sampler and the uniform sampler is large as $323$ and $385$, respectively, which can be a large advantage.
However, when compared to our quantum-inspired classical algorithm, the ratio becomes only about $11$ and $15$.
It implies that we only need about $\sim 10$ times more samples to obtain the instance.

\begin{figure}[t]
\includegraphics[width=250px]{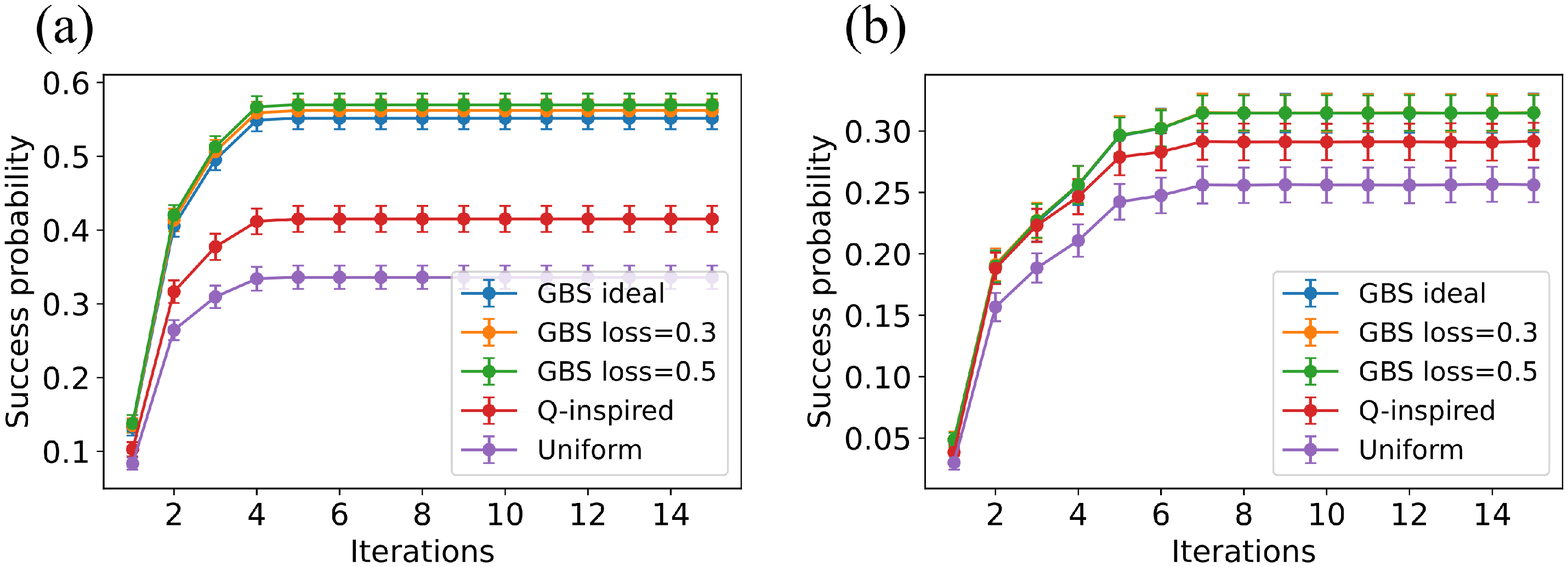} 
\caption{Success probability of finding the maximum weighted clique with different iteration steps for shrinking and searching with the graphs (a) TACE-AS and (b) 1ow7, used in Ref.~\cite{banchi2020molecular}.}
\label{fig:clique}
\end{figure}


After sampling from each sampler, we have also implemented additional heuristic procedures, so-called shrinking and searching \cite{pullan2006dynamic, pullan2006phased}.
The basic idea is to shrink the sampled subgraphs to a smaller clique by truncating vertices that do not constitute a clique and to locally search larger cliques by adding vertices that compose cliques.
After the procedures, we compare the success probability of finding the maximum weighted cliques, which is shown in Fig.~\ref{fig:clique}.
One can see that after the additional step, the difference is not as dramatic as in the first step and that our quantum-inspired classical sampler performs better than the uniform sampler, which has frequently been used as a benchmark.

Now we compare the maximum squeezing parameters required for ideal and lossy Gaussian boson samplers.
For ideal case, for the first graph, the maximum squeezing parameter is $r_\text{max}=1.380$.
On the other hand, for the lossy case, when the loss rate is $1-\eta=0.3$, $r_\text{max}=1.557$ and when the loss rate is $1-\eta=0.5$, $r_\text{max}=1.725$.
For the second graph, for the ideal case and lossy cases with $1-\eta=0.3,0.5$, the maximum squeezing parameters are given by $r_\text{max}=1.121, 1.312, 1.499$, respectively.
Therefore, although theoretically, the lossy case is comparable or even better than the lossless case, the initial squeezing parameters to compensate for the realistic photon loss become more demanding in experiments.

\section{Discussion on potential quantum advantage from Gaussian boson sampler}\label{sec:advantage}
Finally, we discuss how much improvement a Gaussian boson sampler might achieve over our classical algorithm and if a Gaussian boson sampler can provide an exponential speed-up over our algorithm.
We emphasize that finding the densest $k$-subgraph or the maximum weighted clique relies on the heuristic argument for the proportionality between the number of perfect matchings and the density.
To avoid the subtlety due to this, in this section, we will focus on the problem that finds the subgraphs with the maximum number of perfect 
matchings, which is directly related to the output probabilities \cite{arrazola2018quantum}.
To do this, we consider the problem of obtaining a specific sample $\bm{n}^*$ having the largest number of perfect matchings, say the solution of the problem, and compare how much time cost is required for each sampler with fixing a certain photon number sector.

Recall that for a given adjacency matrix $A$, a Gaussian boson sampler generates samples following the probability distribution (normalized within a particular photon number sector)
\begin{align}
    p_Q(\bm{n})=\frac{|\haf(A_{\bm{n}})|^2}{Z_Q}=\frac{\haf(A_{\bm{n}})^2}{Z_Q},
\end{align}
and the proposed quantum-inspired classical algorithm generates samples following the probability
\begin{align}
    p_C(\bm{n})=\frac{\haf(A_{\bm{n}})}{Z_C}.
\end{align}
Here, $Z_Q$ and $Z_C$ are normalization factors.
Then the two distributions are related as
\begin{align}
    p_C(\bm{n})=\frac{\sqrt{p_Q(\bm{n})}}{\sum_{\bm{n}}\sqrt{p_Q(\bm{n})}}.
\end{align}

First, we assume that the solution $\bm{n}^*$ is unique, which can be generalized to polynomially many solutions.
In this case, when $p_Q(\bm{n}^*)$ is inverse-exponentially small, the Gaussian boson sampler already takes exponential time to obtain the corresponding sample.
Thus, we will focus on the case that $p_Q(\bm{n}^*)$ is only inverse-polynomially small, which guarantees that a Gaussian boson sampler can generate the corresponding sample in a polynomial time.
Now, suppose that the probabilities $p_Q(\bm{n})$ are concentrated on polynomially many outcomes including $\bm{n}^*$, i.e., $p_Q(\bm{n})=O(1/\text{poly}(M))$ for some polynomially many $\bm{n}$ and $p_Q(\bm{n})=0$ otherwise.
Then, using the Cauchy-Schwarz inequality,
\begin{align}
    \sum_{\bm{n}}\sqrt{p_Q(\bm{n})}
    \leq \sqrt{\sum_{\bm{n}}p_Q(\bm{n})\sum_{\bm{n}}1^2}
    &=\sqrt{\sum_{\bm{n}}1^2} \nonumber \\ 
    &=O(\text{poly}(M)),
\end{align}
where the sum is over the support of $p_Q(\bm{n})$,
we obtain for $\bm{n}$ such that $p_Q(\bm{n})=O(1/\text{poly}(M))$, $p_C(\bm{n})$ is at least inverse-polynomially large.
Therefore, the classical sampler also takes only polynomial time to obtain the sample $\bm{n}^*$, which implies that the Gaussian boson sampler provides at most a polynomial speed-up.

Now, suppose that $p_Q(\bm{n})=O(1/\text{poly}(M))$ for polynomially many $\bm{n}$ including $\bm{n}^*$ and $p_Q(\bm{n})=O(1/\text{exp}(M))$ for other exponentially many outcomes.
For simplicity, let us consider an example: $p_Q(\bm{n}^*)=3/4$ and $p_Q(\bm{n})=1/4^{M+1}$ for $4^M$ number of $\bm{n}$.
For such a distribution, the corresponding probability $p_C(\bm{n}^*)$ becomes
\begin{align}
    p_C(\bm{n}^*)
    =\frac{\sqrt{3}/2}{\sqrt{3}/2+1/2^{M+1}\times 4^M}
    =O(1/\exp(M)),
\end{align}
which is inverse-exponentially small.
Therefore, for this example, running the quantum-inspired classical algorithm we proposed would take an exponential time to generate $\bm{n}^*$, which shows an exponential separation between the Gaussian boson sampling and our quantum-inspired classical algorithm.

The pertinent question is whether such a nontrivial problem with the above distribution exists.
Unfortunately, it might not be the case.
Let us consider a similar problem in the Fock-state boson sampling setting \cite{aaronson2011computational}, where the probability is expressed by the permanent of the submatrices of a unitary matrix.
Mathematically, it was shown that if the output probability (permanent) is large as $1/\poly$, the relevant matrix has to be very close to a trivial matrix, such as the identity matrix, or permutation matrices \cite{aaronson2014near, berkowitz2018stability}.
We expect that a similar result holds for the hafnian and the Gaussian boson sampling, although it has not been proven to the best of our knowledge. (If it turns out to be false for hafnian, it can provide an exponential speed-up over our algorithm, as shown above.)
In addition, since the relevant graph-theoretic problems, such as finding the densest $k$-subgraphs and finding the maximum weight clique, are NP-hard, such instances are not the hardest instances of the problems unless quantum devices can solve NP-hard problems efficiently.
Nevertheless, the above example implies that if a relevant problem associated with an adjacency matrix $A$ has an outcome $\bm{n}^*$ with large $|\haf(A_{\bm{n}^*})|^2$ with other exponentially many outcomes having very small probabilities, running a Gaussian boson sampler can provide a significant improvement over our algorithm.

Another more interesting possibility for quantum advantage is the problem in which there exist exponentially many solutions among a much larger number of possible instances, instead of a unique solution.
Suppose that we have $4^N/2$ number of solutions that have probability $p_Q(\bm{n})=1/4^N$ and $4^{M}/2$ number of instances that are not solutions whose probabilities are $4^{M}$, where $N$ is a positive integer.
Then, the probability of obtaining a solution is given by $1/2$ for the quantum case.
If we translate this into our classical algorithm, the probability of obtaining a solution is given by
\begin{align}
    \sum_{\bm{n}:\text{sol}}p_C(\bm{n})
    =\frac{\sum_{\bm{n}:\text{sol}}\sqrt{p_Q(\bm{n})}}{\sum_{\bm{n}}\sqrt{p_Q(\bm{n})}}
    =\frac{2^{N}}{2^N+2^M}
    =\frac{1}{1+2^{M-N}}.
\end{align}
Hence, when $M-N=\Omega(N)$, e.g., $M=2N$, the probability of obtaining a solution becomes exponentially small. 
Therefore, in this case, whereas the Gaussian boson sampler can efficiently obtain a solution, our classical algorithm takes exponential samples, $2^{M-N}$, to obtain a solution.
Unlike the unique-solution case, the Gaussian boson sampler may still provide an advantage over our algorithm.

\section{Discussions}
In this work, we have presented a quantum-inspired classical algorithm for finding the densest $k$-subgraph and finding the maximum weighted clique.
We numerically show that although the Gaussian boson sampler may provide an advantage over our algorithm for a fixed number of samples, the advantage is generally not very significant.
On the other hand, we provide a potential advantage of a Gaussian boson sampler over our classical algorithm  in Sec.~\ref{sec:advantage}.

However, it might be possible to find a better classical algorithm than our algorithm; in principle, a classical algorithm following the same probability distribution as the Gaussian boson sampler associated with a nonnegative matrix $A$ is not prohibited because computing the hafnian of a nonnegative matrix within a multiplicative error is in $\text{BPP}^\text{NP}$ not \#P-hard, which is crucial for hardness proof of boson sampling \cite{stockmeyer1983complexity, aaronson2011computational}.
Furthermore, since finding the densest $k$-subgraph and the maximum weighted clique relies on the proportionality between the density and the number of perfect matchings, there may exist a classical sampler whose output probability is proportional to a quantity that is easy to sample from but is still highly proportional to the density.
Again, we emphasize that our algorithm is irrelevant to the standard Gaussian boson sampling hardness \cite{hamilton2017gaussian, deshpande2021quantum} because the associated matrix for the standard setting is not generally nonnegative.
In addition, analyzing and generalizing our algorithm for other applications \cite{bromley2020applications, mezher2023solving}, such as optimization \cite{arrazola2021quantum} and graph similarity \cite{bradler2021graph, schuld2020measuring, bradler2019duality, sempere2022experimentally}, would be an interesting future work.


In Ref.~\cite{bulmer2021boundary}, the so-called Independent Pairs and Singles (IPS) distribution, which was developed for a different purpose, is shown to be classically efficiently samplable.
For our purpose, we focus only on Pairs.
For a given $M\times M$ nonnegative matrix $A$, i.e., $A_{jk}\geq 0$, the distribution is defined as follows:
\begin{align}
    Q(\bm{n})
    =\frac{e^{-\frac{1}{2}\sum_{j,k=1}^MA_{jk}}}{\prod_i n_i!}\haf(A_{\bm{n}}),
\end{align}
where $\bm{n}$ corresponds to the output photon number vector.
The sampling algorithm is implemented by generating photon pairs for each mode pairs $(j,k)$ (with $j\leq k$) from a Poisson distribution with mean given by $A_{j,k}$ and combining the photon numbers.
First, if we post-select collision-free outcomes for a fixed total photon number, then the probability distribution is equivalent to our proposed algorithm.
However, for a given matrix $A$, the total photon number distribution of the IPS sampler is fixed, which potentially costs an additional overhead. (This is the case for Gaussian boson samplers as well.)
More specifically, when one wants to focus on subgraphs with a fixed number of vertices, the post-selection overhead is additionally required.
On the other hand, our algorithm can generate the desired photon number only, which does not cause any post-selection cost (except for collision-free post-selection, which applies to Gaussian boson samplers and IPS samplers as well.).

\appendix

\begin{acknowledgements}
We thank L. Banchi, M. Fingerhuth, T. Babej, C. Ing, and J. M. Arrazola for providing the information of graphs from Ref.~\cite{banchi2020molecular}.
We acknowledge support from the ARO MURI (W911NF-16-1-0349, W911NF-21-1-0325), AFOSR MURI (FA9550-19-1-0399, FA9550-21-1-0209), DoE Q-NEXT, NSF (OMA-1936118, ERC-1941583, OMA-2137642), NTT Research, and the Packard Foundation (2020-71479).
N.Q. acknowledges support from the Minist\`{e}re de l'\'{E}conomie et de l’Innovation du Qu\`{e}bec and the Natural Sciences and Engineering Research Council of Canada.
The authors are also grateful for the support of the University of Chicago Research Computing Center for assistance with the numerical simulations carried out in this paper.
We acknowledge The Walrus python library for the open source of Gaussian boson sampling algorithms \cite{gupt2019walrus}.
\end{acknowledgements}

\bibliography{reference.bib}

\end{document}